\documentclass[reprint,aps,prb,superscriptaddress]{revtex4-1}
\pdfoutput=1

\usepackage{bm}
\usepackage{amsmath,amsthm,amssymb}
\usepackage{array,booktabs,longtable}
\usepackage{mathrsfs}
\usepackage{subfigure}
\usepackage{color}
\usepackage{tikz}
\usetikzlibrary{calc}
\usepackage{array}

\newcommand{\bra}[1]{\langle #1 |}
\newcommand{\ket}[1]{| #1\rangle}

\newcommand{\bmm}{\begin{matrix}}
\newcommand{\emm}{\end{matrix}}
\newcommand{\BLvert}{\Biggl\vert\bmm} 
\newcommand{\Brangle}{\emm\Biggr\rangle}

\makeatletter
\newcommand{\thickhline}{%
    \noalign {\ifnum 0=`}\fi \hrule height 1.25pt
    \futurelet \reserved@a \@xhline
}
\newcolumntype{"}{@{\hskip\tabcolsep\vrule width 1.25pt\hskip\tabcolsep}}
\makeatother

\newcommand{\oneTriangle}[4][0]{
\begin{tikzpicture}[scale=1,baseline]
\draw (0,0) node[left] {\scalebox{0.7}{$#2$}} -- (1,0) node[right] {\scalebox{0.7}{$#3$}} -- (60:1) node[above] {\scalebox{0.7}{$#4$}} -- cycle;
\ifnum #1=1 {
   \draw[-<,>=latex,line width=0.01pt] (0,0) -- (0.5,0) node[below] {\scalebox{0.7}{$[#2#3]$}};
   \draw[-<,>=latex,line width=0.01pt] (0,0) -- (60:0.5) node[left] {\scalebox{0.7}{$[#3#4]$}};
   \draw[->,>=latex,line width=0.01pt] (0,0) +(60:1) -- +(30:0.866) node[right] {\scalebox{0.7}{$[#2#4]$}};}
\else {
   \draw[-<,>=latex,line width=0.01pt] (0,0) -- (0.5,0) ;
   \draw[-<,>=latex,line width=0.01pt] (0,0) -- (60:0.5) ;
   \draw[->,>=latex,line width=0.01pt] (0,0) +(60:1) -- +(30:0.866);}  \fi
\end{tikzpicture}
}

\newcommand{\oneSquare}[6]{
\begin{tikzpicture}[scale=1,baseline]
\draw (0,0) -- (1,0)  -- (1,1) -- (0,1) -- cycle;

   \draw[->,>=latex,line width=0.01pt] (0,0) -- (0.5,0) node[below] {\scalebox{0.7}{$#4$}};
   \draw[->,>=latex,line width=0.01pt] (0,0) -- (0,0.5) node[left] {\scalebox{0.7}{$#1$}};
   \draw[->,>=latex,line width=0.01pt] (0,1) -- (0.5,1)	node[above] {\scalebox{0.7}{$#2$}};
   \draw[->,>=latex,line width=0.01pt] (1,0) -- (1,0.5)	node[right] {\scalebox{0.7}{$#3$}};
   
   \draw (0.5,0.5) node {\scalebox{0.7}{$#5$}};
   
\ifnum #6=0 {
	   \draw[->] (0.65,0.4) arc (320:0:0.2cm);
}
\else{
	\draw[->] (0.65,0.5) arc (0:320:0.2cm);
} \fi
   
\end{tikzpicture}
}

\newcommand{\oneVertex}[6]{
\begin{tikzpicture}[scale=0.75,baseline]
\draw (0,-1) node[right] {\scalebox{0.7}{$#5$}} -- (0,1) node[left] {\scalebox{0.7}{$#2$}};
\draw (-1,0) node[above] {\scalebox{0.7}{$#1$}} -- (1,0) node[above] {\scalebox{0.7}{$#4$}};
\draw (0,0) +(52:1) node[right] {\scalebox{0.7}{$#3$}} -- (0,0);
\draw (0,0) +(232:1) node[left] {\scalebox{0.7}{$#6$}} -- (0,0);

\draw[->,>=latex,line width=0.01pt] (0,0) -- (0.5,0);
\draw[->,>=latex,line width=0.01pt] (0,0) -- (0,0.5);
\draw[->,>=latex,line width=0.01pt] (-1,0) -- (-0.5,0);
\draw[->,>=latex,line width=0.01pt] (0,-1) -- (0,-0.5);
\draw[-<,>=latex,line width=0.01pt] (0,0) -- ($(0,0) +(232:0.5)$);
\draw[->,>=latex,line width=0.01pt] (0,0) -- ($(0,0) +(52:0.5)$);

\draw (-0.3,0) node[above] {\scalebox{0.7}{$s$}};
\end{tikzpicture}
}

\begin{document}


\title{Universal Topological Data for Gapped Quantum Liquids in Three Dimensions\\
and Fusion Algebra for Non-Abelian String Excitations}



\author{Heidar Moradi}
\affiliation{Perimeter Institute for Theoretical Physics, Waterloo, Ontario, N2L 2Y5 Canada}

\author{Xiao-Gang Wen}
\affiliation{Perimeter Institute for Theoretical Physics, Waterloo, Ontario, N2L 2Y5 Canada} 
\affiliation{Department of Physics, Massachusetts Institute of Technology, Cambridge, Massachusetts 02139, USA}

\date{\today}

\begin{abstract}
Recently we conjectured that a certain set of universal topological quantities
characterize topological order in any dimension.  Those quantities can be
extracted from the universal overlap of the ground state wave functions.  For
systems with gapped boundaries, these quantities are representations of the
mapping class group $\texttt{MCG}(\mathcal M)$ of the space manifold $\mathcal
M$ on which the systems lives. We will here consider simple examples in three
dimensions and give physical interpretation of these quantities, related to
fusion algebra and statistics of particle and string excitations. In
particular, we will consider dimensional reduction from 3+1D to 2+1D, and show
how the induced 2+1D topological data contains information on the fusion and
the braiding of non-Abelian string excitations in 3D.  These universal
quantities generalize the well-known modular $S$ and $T$ matrices to any
dimension.

\end{abstract}


\maketitle


\section{Introduction}

For more than two decades exotic quantum
states\cite{KDP8094,TSG8259,L8395,KL8795,WWZ8913,RS9173,W9164,MS0181,MR9162,W9102,WES8776,RMM0899}
have attracted a lot attention from the condensed matter community.
In particular gapped systems with non-trivial topological
order,\cite{Wtop,WNtop,Wrig} which is a reflection of long-range
entanglement\cite{CGW1038} of the ground state, have been studied intensely in
$2+1$ dimensions. Recently, people started to work on a general theory of
topological order in higher than $2+1$
dimensions.\cite{LWstrnet,WW1132,MW14,WL1437,JMR1462}

In a recent work \Ref{MW14}, we conjectured that for a gapped system on a
$d$-dimensional manifold $\mathcal M$ of volume $V$ with the set of degenerate
ground states $\{\ket{\psi_\alpha}\}_{\alpha=1}^N$ on $\mathcal M$, we have the
following overlaps
\begin{equation}\label{eq:overlap}
	\bra{\psi_\alpha}\hat{\mathcal O}_A\ket{\psi_\beta} = e^{-\alpha V + o(1/V)}M_{\alpha,\beta}^A,
\end{equation}
where $\hat{\mathcal O}_A$ are transformations on the wave functions induced by
the automorphisms $A:\mathcal M\rightarrow\mathcal M$, $\alpha$ is a
non-universal constant and $M^A$ is a universal matrix up to an overall $U(1)$
phase. Here $M^A$ form a projective representation of the automorphism group
$\texttt{AMG}(\mathcal M)$, which is robust against \textit{any} local
perturbations that do not close the bulk gap.\cite{Wrig,KW9327} In \Ref{MW14}
we conjectured that such projective representations for different space
manifold topologies fully characterize topological orders with finite ground
state degeneracy in any dimension. Furthermore, we conjectured that projective
representations of the mapping class groups $\texttt{MCG}(\mathcal M) =
\pi_0[\texttt{AMG}(\mathcal M)]$ classify topological order with gapped
boundaries.\cite{Wrig,KW9327} These quantities can be used as order parameters for topological order and detect transitions between different phases.\cite{HMW1457}

In this paper we will study these universal quantities further in 3-dimensions
for one of the most simple manifolds, the 3-torus $\mathcal M=T^3$. The mapping
class group of the 3-torus is $\texttt{MCG}(T^3) = SL(3,\mathbb Z)$. This group
is generated by two elements of the form \cite{trott1962pair}
\begin{equation}\label{eq:SL3ZGenerators}
	\hat{\tilde S} = \begin{pmatrix}
				0 & 1 & 0 \\
				0 & 0 & 1 \\
				1& 0 & 0
			 \end{pmatrix},
			 \qquad
	\hat{\tilde T}= \begin{pmatrix}
				1 & 0 & 0 \\
				1 & 1 & 0 \\
				0 & 0 & 1
			 \end{pmatrix}.
\end{equation}
These matrices act on the unit vectors by $\hat{\tilde S}:(\hat{\bm x},
\hat{\bm y} , \hat{\bm z}) \mapsto (\hat{\bm z}, \hat{\bm x} , \hat{\bm y})$
and similarly $\hat{\tilde T}:(\hat{\bm x}, \hat{\bm y} , \hat{\bm z}) \mapsto
(\hat{\bm x} + \hat{\bm y}, \hat{\bm y} , \hat{\bm z})$. Thus $\tilde S$
corresponds to a rotation, while $\tilde T$ is shear transformation in the
$xy$-plane.

In this paper, we will study the $SL(3,\mathbb Z)$ representations generated by
a very simple class of $\mathbb Z_N$ models in detail and then consider models
for any finite group $G$, which are 3-dimensional versions of Kitaevs quantum
double models \cite{K032}. One can also generalize into twisted
versions of these based on the group cohomology $H^4(G,U(1))$ by
direct generalization of \Ref{HWW1314} into 3+1D.\cite{JMR1462,WW14}

We will consider dimensional reduction of a 3D topological order $\cC^{3D}$ to
2D by making one direction of the 3D space into a small circle.  In this limit,
the 3D topologically ordered states $\cC^{3D}$ can be viewed as several 2D
topological orders $\cC^{2D}_i$, $i=1,2,\cdots$ which happen to have degenerate
ground state energy.  We denote such a dimensional reduction process as
\begin{equation}
\label{C3C2}
 \cC^{3D} = \bigoplus_i \cC^{2D}_i.
\end{equation}
We can compute such a dimensional reduction using the representation of
$SL(3,\mathbb Z)$ that we have calculated.

We consider $SL(2,\mathbb Z)\subset SL(3,\mathbb Z)$ subgroup and the reduction
of the $SL(3,\mathbb Z)$ representation $R^{3D}$ to the $SL(2,\mathbb Z)$
representations $R^{2D}_i$:
\begin{equation}
\label{R3R2}
 R^{3D} = \bigoplus_i R^{2D}_i.
\end{equation}
We will refer to this as branching rules for the $SL(2,\mathbb Z)$ subgroup.
The  $SL(3,\mathbb Z)$ representation $R^{3D}$ describes the 3D topological
order $\cC^{3D}$ and the  $SL(2,\mathbb Z)$ representations $R^{2D}_i$ describe
the 2D topological orders $\cC^{2D}_i$.  The decomposition \eq{R3R2} gives us the dimensional reduction \eq{C3C2}.

Let us use $\cC_G$ to denote the topological order described by the gauge
theory with the finite gauge group $G$. Using the above result, we find that
\begin{align}
\cC_G^{3D} = \bigoplus_{n=1}^{{|G|}}\cC_G^{2D} 
\end{align}
for Abelian $G$ where $|G|$ is the number of the group elements.
For non-Abelian group $G$
\begin{align}
\label{G3G2}
\cC_G^{3D} = \bigoplus_{C}\cC_{G_C}^{2D} 
\end{align}
where $\bigoplus_{C}$ sums over all different conjugacy classes $C$ of $G$, and
$G_C$ is a subgroup of $G$ which commutes with an element in $C$.  The results
for $G=\mathbb Z_N$ were mentioned in our previous paper.\cite{MW14}

We also found that the reduction of $SL(3,\mathbb Z)$ representation,
\eqn{R3R2}, encodes all the information about the three-string statistics
discussed in \Ref{WL1437,JMR1462} for Abelian groups.  For non-Abelian groups,
we will have a ``non-Abelian'' string braiding statistics and a non-trivial
string fusion algebra.  We also have  a ``non-Abelian'' three-string braiding
statistics and a non-trivial three-string fusion algebra.  Within the dimension
reduction picture, the 3D strings reduces to particles in 2D, and the
(non-Abelian) statistics of the particles encode the (non-Abelian) statistics
of the strings.

\section{$\mathbb Z_N$ Model in 3-Dimensions}

In this section we will define and study the excitations of a $\mathbb Z_N$
model in detail\footnote{Two-dimensional version of this model has previously
been studied in for example \Ref{SchulzetAlZn}.} and compute the 3-torus
universal matrices, eq. \eqref{eq:overlap}. 

\begin{figure}[tb]
\centering
\subfigure[\ \label{fig:3dVertex}]{
\includegraphics[width=.15\textwidth]{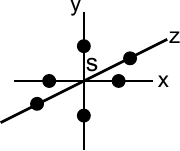}}\qquad
\subfigure[\ \label{fig:3dUnitCell} ]{
\includegraphics[width=.15\textwidth]{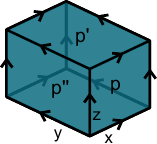}}
\caption{(a) Lattice site of 3D cubic lattice. $A_s$ act on spins connected to site $s$. (b) 2D plaquettes. $B_p$ acts on the four spins surrounding $p$. Choose a righthanded $(x,y,z)$ frame, and let all links be oriented wrt. to these directions. This associates a natural orientation to $2D$ plaquettes on the dual lattice.}\label{fig:3DToricCodeOperators}
\end{figure}

Consider a simple cubic lattice with a local Hilbert space on each link
isomorphic to the group algebra of $\mathbb Z_N$, $\mathcal H_i\approx\mathbb C[\mathbb
Z_N]\approx\mathbb C^N \approx \text{span}_{\mathbb
C}\{\ket{\sigma}|\sigma\in\mathbb Z_N\}$. Give the links on the lattice an
orientation as in figure \ref{fig:3DToricCodeOperators} and let there be a
natural isomorphism $\mathcal H_i\overset{\sim}{\rightarrow}\mathcal
H_{i^\star}$ for link $i$ and its reversed orientation $i^\star$ as
$\ket{\sigma_i}\mapsto\ket{\sigma_{i^\star}}=\ket{-\sigma_i}$. Let this basis
be orthonormal. Define two local operators \[	Z_i\ket{\sigma_i}
=\omega^{\sigma_i}\ket{\sigma_i},\qquad X_i\ket{\sigma_i}=\ket{\sigma_i-1},
\] where $\omega = e^{\frac{2\pi i}N}$.  These operators have the important
commutation relation $X_iZ_i = \omega Z_iX_i$.  Note that these operators are
unitary and satisfy $X_i^N=Z_i^N=1$. For each lattice site $s$ and plaquette
$p$ define \[ A_s = \prod_{i\in s_+}Z_i\prod_{j\in s_-}Z_j^\dagger,\quad B_p =
\prod_{i\in\partial p_+}X_i^\dagger\prod_{j\in\partial p_-}X_j. \] Here $s_+$
is the set of links pointing into $s$, while $s_-$ is the set of links pointing away from $s$. $B_p$ creates a string around
plaquette $p$ with orientation given by the normal direction using the right
hand thumb rule. Then $\partial p_\pm$ are the set of links surrounding
plaquette $p$ with the same or opposite orientation as the lattice. One can
directly check that all these operators commute for all sites and plaquettes.

We can now define the $\mathbb Z_N$ model by the Hamiltonian
\[ H_{3D,\mathbb Z_N} = -\frac{J_e}2\sum_s\left(A_s+A_s^\dagger\right) - \frac{J_m}2 \sum_p\left(B_p + B_p^\dagger\right), \]
where we will assume $J_e, J_m\geq 0$ throughout. Since $\text{eigen}(A_s+A_s^\dagger) = \{2\cos(\frac{2\pi}Nq)\}_0^{N-1}$, and the similar for $B_p+B_p^\dagger$, the ground state is the state satisfying
\begin{equation}\label{eq:GSConditions}
	A_s\ket{GS} = \ket{GS}, \qquad B_p\ket{GS}=\ket{GS},
\end{equation}
for all $s$ and $p$.
We can easily construct hermitian projectors to the state with eigenvalue $1$ for all vertices and plaquettes
\[	\rho_s = \frac 1N\sum_{k=0}^{N-1}A_s^k,\qquad \rho_p = \frac 1N\sum_{k=0}^{N-1}B_p^k.\]
The ground state is thus $\ket{GS} = \prod_s\rho_s\prod_p\rho_p\ket{\psi}$, for any reference state $\ket{\psi}$ such that $\ket{GS}$ is non-zero. For the choice $\ket{\psi} = \ket{00\dots 0}\equiv\ket 0$, the $\rho_s$ is trivial and the ground state is thus
\[	\ket{GS} = \prod_p\left(\frac 1N\sum_{k=0}^{N-1}B_p^k\right)\ket 0 = \mathcal N\sum_{\mathbb Z_N \\ \text{ string nets}}\ket{\text{loops}}. \]
The first condition in equation \eqref{eq:GSConditions} requires that the ground state consists of $\mathbb Z_N$ string-nets, while the second requires that these appear with equal superpositions. Note that if we had used eigenstates of $X_i$ instead, we would find that the ground state is a membrane condensate on the dual lattice.

\subsubsection{String and Membrane Operators}

Now let $l_{ab}$ denote a curve on the lattice from site $a$ to $b$, with the orientation that it points from $a$ to $b$. And let $\Sigma_{\mathcal C}$ denote an oriented surface on the dual lattice with $\partial\Sigma_{\mathcal C} = \mathcal C$. Using these, define string and membrane operators
\[ W[l_{ab}]=\prod_{i\in l^-_{ab}}X_i\prod_{j\in l^+_{ab}}X_j^\dagger, \quad \Gamma[\Sigma_{\mathcal C}]=\prod_{i\in\Sigma^-_{\mathcal C}}Z_i^\dagger\prod_{j\in\Sigma^+_{\mathcal C}}Z_j.	\]
Again $l^\pm_{ab}$ and $\Sigma_{\mathcal C}^\pm$ are defined wrt. the orientation of the lattice.
Note that $B_p = W[\partial p]$, where $\partial p$ denotes a closed loop around plaquette $p$ with right hand thumb rule orientation wrt. the normal direction. Similarly, $A_s = \Gamma[\text{star}(s)]$, where $\text{star}(s)$ is the closed surface on the dual lattice surrounding site $s$ with inward orientation.

It is clear that the following operators commute
\[	\Big[W[l_{ab}],B_p\Big]=0, \quad \forall p,\quad\text{and}\quad \Big[\Gamma[\Sigma_{\mathcal C}],A_s\Big]=0, \quad \forall s. 	\]
Furthermore it is easy to show that
\[	\Big[W[l_{ab}],A_s\Big]=0,\quad s\neq a,b,\qquad  \Big[\Gamma[\Sigma_{\mathcal C}],B_p\Big]=0,\quad p\not\in\mathcal C,	\]
while
\[	A_aW[l_{ab}] = \omega^{-1}\:W[l_{ab}]A_a, \qquad A_bW[l_{ab}] = \omega \:W[l_{ab}]A_b,	\]
and
\[	B_p \Gamma[\Sigma_{\mathcal C}] = \omega^{\pm 1}\: \Gamma[\Sigma_{\mathcal C}]B_p, \quad p\in\mathcal C,	\]
where $\pm$ depends on orientation of $\Sigma_{\mathcal C}$.
\subsubsection{Ground States on 3-Torus}
The ground state degeneracy depends on the topology of the manifold on which the theory is defined, take for example the 3-torus $T^3$.
Let $l_x$, $l_y$ and $l_z$ be non-contractible loops along the three cycles on the lattice, with the orientation of the lattice. Similarly, let $\Sigma_x$, $\Sigma_y$ and $\Sigma_z$ be non-contractible surfaces along the three-directions, with the orientation of the dual lattice. We can define the operators
\[ W_i\equiv W[l_i] = \prod_{j\in l_i}X^\dagger_j, \quad \Gamma_i\equiv\Gamma[\Sigma_i] = \prod_{j\in\Sigma_i}Z_i,\quad i=x,y,z.	\]
These operators have the commutation relations
\begin{equation}
	W_i\Gamma_i = \omega^{-1}\:\Gamma_i W_i,\qquad i=x,y,z.
\end{equation}

We can thus find three commuting (independent) non-contractible operators to get $N^3$ fold ground state degeneracy. For example $	\ket{\alpha,\beta,\gamma} = (W_x)^\alpha(W_y)^\beta(W_z)^\gamma\ket{GS}$,
where $\alpha,\beta,\gamma = 0,\dots, N-1$. This basis correspond to eigenstates of the surface operators $\Gamma_i\ket{\alpha_1,\alpha_2,\alpha_3} = \omega^{\alpha_i}\ket{\alpha_1,\alpha_2,\alpha_3}$.
Note that on the torus we get the extra set of constraints $\prod_sA_s = 1$, $\prod_pB_p =1$.
Let $G$ be the group generated by $B_p$ for all $p$, modulo $B_pB_{p'}=B_{p'}B_p$, $B_p^N=1$ and $\prod_pB_p=1$. Furthermore define the groups $G_{\alpha\beta\gamma} \equiv (W_x)^\alpha(W_y)^\beta(W_z)^\gamma G$, then we can write the ground states as
\[ \ket{\alpha,\beta,\gamma} = \frac 1{\sqrt{|G_{\alpha\beta\gamma}|}}\sum_{g\in G_{\alpha\beta\gamma}}\ket{g},\]
where $\ket g\equiv g\ket 0$.

In 2D, the quasiparticle basis corresponds to the basis in which there is well-defined magnetic and electric flux along one cycle of the torus. We can try to do the same in three-dimensions. $\Gamma_x$, $W_y$, $W_z$ all commute with each other and we can consider the basis which diagonalizes all of them.
This basis is given by
\begin{equation}\label{eq:3DZNQuasiParticleBasis}
	\ket{\psi_{abc}} = \frac 1N\sum_{\beta\gamma}\omega^{-\beta b-\gamma c}\ket{a,\beta,\gamma},
\end{equation}
where $a,b,c=0,\dots, N-1$.
These are clearly eigenstates of $\Gamma_x$, and furthermore we have that $W_y\ket{\psi_{abc}}=\omega^b\ket{\psi_{abc}}$ and $W_z\ket{\psi_{abc}}=\omega^c\ket{\psi_{abc}}$.
This basis is a 3D version of minimum entropy states (MES).\cite{ZGT1251}

\subsubsection{Excitations}
Now lets go back to, say, this theory on $S^3$ and look at elementary excitations of our model. An excitation correspond to a state in which the conditions \eqref{eq:GSConditions} are violated in a small region.
Using the string operators, we can create a pair of particles by $\ket{-q_e,q_e} = W[l_{ab}]^{q_e}\ket{GS}$
with the electric charges
\[	A_a\ket{-q_e,q_e} = \omega^{-q_e}\ket{-q_e,q_e}, \quad A_b\ket{-q_e,q_e} = \omega^{q_e}\ket{-q_e,q_e}.	\]
This excitation has an energy cost of $\Delta E_{\text{particles}} = 2J_e[1-\cos(\frac{2\pi}Nq_e)]$.
Furthermore we have oriented string excitations by using the membrane operators $\ket{\mathcal C, q_m} = \Gamma[\Sigma_{\mathcal C}]^{q_m}\ket{GS}$, with the magnetic flux
\[	B_p \ket{\mathcal C, q_m} = \omega^{\pm q_m}\ket{\mathcal C, q_m}, \quad p\in\mathcal C,	\]
where the $\pm$ depend on the orientation of $\mathcal C$.
This excitation comes with the energy penalty $\Delta E_{\text{string}}=\text{Lenght}(\mathcal C)J_m[1-\cos(\frac{2\pi}Nq_m)]$.

\begin{figure}[tb]
\centering
\includegraphics[width=.45\textwidth]{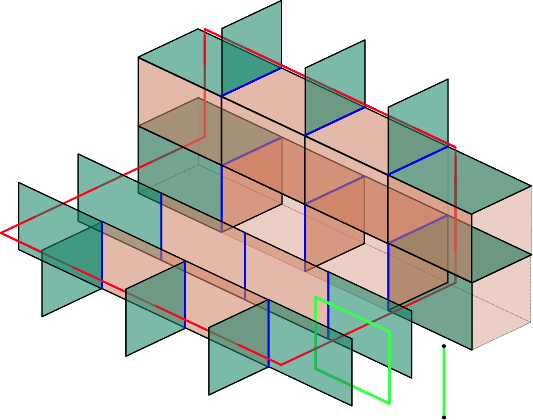}
\caption{String and particle excitations. The red curve is the boundary of a membrane on the dual lattice and correspond to a string excitation. The blue links are the ones affected by the membrane operator and the green plaquettes are the ones on which $B_p$ can measure the presence of the string excitation. The green line correspond to a string operator on the lattice, in which the end point are particles. Mutual statistics between strings and particles can be calculated by creating a particle-antiparticle pair from the vacuum, moving one particle around the string excitation and annihilating the particles.}\label{fig:3DExcitations}
\end{figure}

One can easily show that all the particles have trivial self and mutual statistics, and the same with the strings. Mutual statistics between particles and strings can be non-trivial however, taking a charge $q_e$ particle through a flux $q_m$ string gives the anyonic phase $\omega^{\pm q_eq_m}$, where the $\pm$ depend on the orientations. See figure \ref{fig:3DExcitations}.

\section{Representations of $\texttt{MCG}(T^3)=SL(3,\mathbb Z)$}
Let us now go back to $T^3$ and consider the universal quantities as defined in \eqref{eq:overlap}.
In the $\ket{\alpha,\beta,\gamma}$ basis, the representation of the $SL(3,\mathbb Z)$ generators \eqref{eq:SL3ZGenerators} is given by
\begin{equation}\label{eq:3dStilde}
	\tilde S_{\alpha\beta\gamma,\alpha'\beta'\gamma'} = \delta_{\alpha,\beta'}\delta_{\beta,\gamma'}\delta_{\gamma,\alpha'},
\end{equation}
and
\begin{equation}\label{eq:3dTtilde}
	\tilde T_{\alpha\beta\gamma,\alpha'\beta'\gamma'} = \delta_{\alpha,\alpha'}\delta_{\beta,\alpha'+\beta'}\delta_{\gamma,\gamma'}.
\end{equation}
In the 3D quasiparticle basis \eqref{eq:3DZNQuasiParticleBasis} these are given by
\[
	\tilde S_{abc,\bar a\bar b\bar c} = \frac 1N\delta_{b,\bar c}e^{\frac{2\pi i}N(\bar ac - a\bar b)},\quad \tilde T_{abc,\bar a\bar b\bar c} = \delta_{a,\bar a}\delta_{b,\bar b}\delta_{c,\bar c}e^{\frac{2\pi i}N ab}.
\]
For example in the simplest case $N=2$, which is the 3D Toric code, we have
\[
	\tilde T = \begin{pmatrix}
					1 &&&&&&& \\ & 1 &&&&&& \\ && 1 &&&&& \\ &&& -1 &&&&\\
					&&&& 1 &&& \\ &&&&& 1 && \\ &&&&&& 1 & \\ &&&&&&& -1
		   	   \end{pmatrix},
\]
and
\[
	\tilde S = \frac 12 \begin{pmatrix}
					1 &1&0&0&1&1&0&0 \\ 1& 1 &0&0&-1&-1&0&0 \\ 1&-1& 0 &0&1&-1&0&0 \\ 1&-1&0& 0 &-1&1&0&0\\
					0&0&1&1& 0&0&1&1 \\ 0&0&1&1&0&0 &-1&-1 \\ 0&0&1&-1&0&0& 1 &-1 \\ 0&0&1&-1&0&0&-1& 1
		   	   \end{pmatrix}.
\]
\subsubsection{Interpretation of $\tilde T$}
These matrix elements in this particular ground state basis, actually contain some physical information about statistics of excitations. In order to see this, we can associate a collection of excitations to each ground state on the 3-torus.

First cut the 3-torus along the $x$-axis such that it now has two boundaries.
We can measure the presence of excitations on the boundary using the operators $\Gamma_x$, $W_y$ and $W_z$. First take the state with no particle, $\ket{\bm 1} = \frac 1N\sum_{\beta\gamma}\ket{\beta,\gamma}$, in which all operators have eigenvalue $1$. Here $\ket{\beta,\gamma}$ are states with $\beta$ and $\gamma$ non-contractible electric loops along the $y$ and $z$ axis, respectively.
\begin{figure}[tb]
\centering
\includegraphics[width=.45\textwidth]{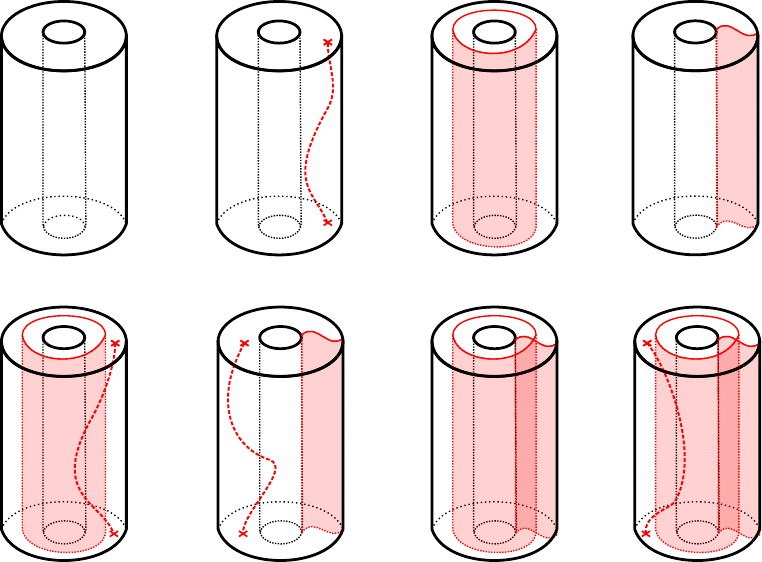}
\caption{The result of cutting open the 3-torus along the $x$-axis, can be represented by a hollow solid cylinder where the inner and outer surfaces are identified, but there are two boundaries along $x$. In the above, the compactified direction is $y$ and the radial direction is $z$, while the open direction is $x$. We can see the $N^3$ possible excitations on the boundaries which give rise to 3-torus ground states uppon gluing. The four first states correspond to $\ket{\bm 1}$, $\ket{e_a}$, $\ket{m_{y,c}}$ and $\ket{m_{z,b}}$. }\label{fig:SolidHollowTorus}
\end{figure}
Now add excitations on the boundary using open string and membrane operators (see fig. \ref{fig:SolidHollowTorus}) $\ket{e_a} =(W[l_{12}])^a\ket{\bm 1}$,
$\ket{m_{y,c}} =(\Gamma[\Sigma_{\mathcal C_y}])^c\ket{\bm 1}$,
$\ket{m_{z,b}} =(\Gamma[\Sigma_{\mathcal C_z}])^b\ket{\bm 1}$,
$\ket{e_am_{y,c}} =(W[l_{12}])^a(\Gamma[\Sigma_{\mathcal C_y}])^c\ket{\bm 1}$,
$\ket{e_am_{z,b}} =(W[l_{12}])^a(\Gamma[\Sigma_{\mathcal C_z}])^b\ket{\bm 1}$,
$\ket{m_{y,c}m_{z,b}} =(\Gamma[\Sigma_{\mathcal C_y}])^c(\Gamma[\Sigma_{\mathcal C_z}])^b\ket{\bm 1}$
and
$\ket{e_a m_{y,c}m_{z,b}} =(W[l_{12}])^a(\Gamma[\Sigma_{\mathcal C_y}])^c(\Gamma[\Sigma_{\mathcal C_z}])^b\ket{\bm 1}$,
where $a,b,c=1,\dots, N-1$. Or more compactly, $\ket{e_a m_{y,c}m_{z,b}}$, where $a,b,c=0,\dots, N-1$.
Here $l_{12}$ is a curve from one edge to the other, $\Sigma_{\mathcal C_y}$ is a membrane between edges wrapping along the $y$-cycle and $\Sigma_{\mathcal C_z}$ is a membrane between edges wrapping along $z$-cycle. All these have the same orientation as the (dual) lattice.
These states have well-defined electric and magnetic flux wrt. $\Gamma_x$, $W_y$ and $W_z$. Here $m_y$ and $m_z$ correspond to the strings on the boundaries, wrapping around the $y$ and $z$ cycles, respectively.

If we now glue the two boundaries together, we see that for each of these excitations we have a 3-torus ground state
\begin{align*}
	\ket{\bm 1} &= \ket{\psi_{000}}, &	\ket{e_am_{1,c}} &= \ket{\psi_{a0c}}, \\ 	\ket{e_a} &= \ket{\psi_{a00}}, &	\ket{e_am_{2,b}} &= \ket{\psi_{ab0}},\\
	\ket{m_{1,c}} &= \ket{\psi_{00c}}, &	\ket{m_{1,c}m_{2,b}} &= \ket{\psi_{0bc}}, \\ 	\ket{m_{2,b}} &= \ket{\psi_{0b0}}, &	\ket{e_am_{1,c}m_{2,b}} &= \ket{\psi_{abc}}.
\end{align*}
We can add other string excitations on the boundary, however they will not give rise to new 3-torus ground states after gluing. We thus see a generalization of the situation in 2D, where there is a direct relation between number of excitation types and GSD on the torus.

Now lets to back to the open boundaries, and consider making a $2\pi$ twist of one of the boundaries, which will give some kind of 3D analogue of \textit{topological spin}. It can be seen that most states will be invariant under such an operation by appropriately deforming and reconnecting the string and membrane operators. For example $\ket{e_a}\rightarrow\ket{e_a}$, which implies that the particles $e_a$ are bosons. However we pick up a factor of $\omega^{ab}$ for $\ket{e_am_{2,b}}$ and $\ket{e_am_{1,c}m_{2,b}}$, since the string corresponding to particle $e_a$ has to cross the membrane corresponding to $m_{2,b}$. Physically this is a consequence of mutual statistics of the particle and string excitation. We can consider these as 3D analogue of topological spin.

\begin{figure}[tb]
\centering
\includegraphics[width=.45\textwidth]{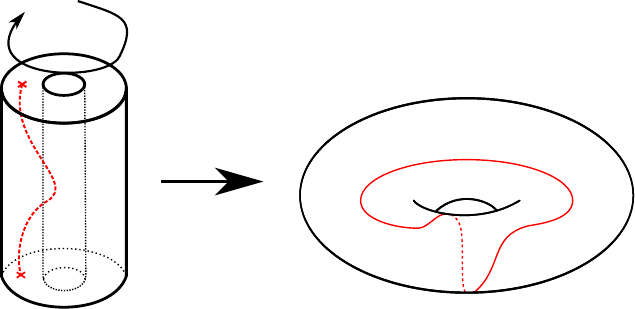}
\caption{The Dehn twist $\tilde T$ is along the $x-y$ plane, thus it is natural to think of $T^3$ as a solid hollow 2-torus where the inner and outer boundaries are identified, here the thickened direction is $z$. In this picture, we can think of $\tilde T$ just as a usual Dehn twist of a 2-torus.}\label{fig:TopologicalSpinDehnTwist}
\end{figure}

Now notice that this operation precisely corresponds to the $\tilde T$ Dehn twist on the 3-torus by gluing the boundaries (see fig.\ref{fig:TopologicalSpinDehnTwist}). Thus $\tilde T$, as calculated from the ground state, should contain information about statistics of excitations. Writing $\tilde T_{abc,\bar a\bar b\bar c} = \delta_{a,\bar a}\delta_{b,\bar b}\delta_{c,\bar c}e^{\frac{2\pi i}N ab}\equiv \delta_{a,\bar a}\delta_{b,\bar b}\delta_{c,\bar c}\tilde T_{abc}$, we get the following 3D \textit{topological spins}
\begin{align*}
	\tilde T_{\bm 1}&= \tilde T_{000} = 1, & \tilde T_{e_a}&= \tilde T_{a00} = 1,\\
	\tilde T_{m_{1,c}}&= \tilde T_{00c} = 1, & \tilde T_{m_{2,b}}&= \tilde T_{0b0} = 1,\\
	\tilde T_{e_am_{1,c}}&= \tilde T_{a0c} = 1, & \tilde T_{e_am_{2,b}}&= \tilde T_{ab0} = e^{\frac{2\pi i}Nab},\\
	\tilde T_{m_{1,c}m_{2,b}}&= \tilde T_{0bc} = 1, & \tilde T_{e_am_{1,c}m_{2,b}}&= \tilde T_{abc} = e^{\frac{2\pi i}Nab}.
\end{align*}
This exactly match the properties of the excitations.
Thus the universal quantity $\tilde T$ calculated from the ground state alone, contain direct physical information about statistics of excitations in the system.
Note that elements like $\tilde T_{m_{1,c}m_{2,b}}$ can be non-trivial in theories with non-trivial string-string statistics.

\subsubsection{$3D\rightarrow 2D$ Dimensional Reduction}
We can actually relate these universal quantities to the well-known $S$ and $T$ matrices in two dimensions.
Consider now the $SL(2,\mathbb Z)$ subgroup of $SL(3,\mathbb Z)$ generated by
\begin{equation}\label{eq:SL2ZsubgroupGenerators}
	\hat T^{yx}\equiv \begin{pmatrix}
								1 & 0 & 0\\
								1 & 1 & 0\\
								0 & 0 & 1
						   \end{pmatrix}
						   \qquad\text{and}\qquad
	\hat S^{yx}\equiv
	 \begin{pmatrix}
								0 & 1 & 0\\
								-1 & 0 & 0\\
								0 & 0 & 1
						   \end{pmatrix}
\end{equation}
One can directly compute the representation of this subgroup for the above $\mathbb Z_N$ model, which is given by
\[
	S^{yx}_{abc,\bar a\bar b\bar c} = \frac 1N\delta_{c,\bar c}e^{-\frac{2\pi i}N(a\bar b + \bar a b)},\quad T^{yx}_{abc,\bar a\bar b\bar c} = \delta_{a,\bar a}\delta_{b,\bar b}\delta_{c,\bar c}e^{\frac{2\pi i}N ab}.
\]
Note that $S^{3D}_{\mathbb Z_N}=\bigoplus_{n=1}^NS^{2D}_{\mathbb Z_N}$ and $T^{3D}_{\mathbb Z_N}=\bigoplus_{n=1}^NT^{2D}_{\mathbb Z_N}$. In particular, for the toric code $N=2$ we have
\[
	S^{yx} = \frac 12 \begin{pmatrix}
					1 &1&1&1&&&& \\ 1& 1 &-1&-1&&&& \\ 1&-1& 1 &-1&&&& \\ 1&-1&-1& 1 &&&&\\
					&&&& 1&1&1&1 \\ &&&&1&1 &-1&-1 \\ &&&&1&-1& 1 &-1 \\ &&&&1&-1&-1& 1
		   	   \end{pmatrix},
\]
and
\[
	T^{yx} = \begin{pmatrix}
					1 &&&&&&& \\ & 1 &&&&&& \\ && 1 &&&&& \\ &&& -1 &&&&\\
					&&&& 1 &&& \\ &&&&& 1 && \\ &&&&&& 1 & \\ &&&&&&& -1
		   	   \end{pmatrix}.
\]
These $N$ blocks are distinguished by eigenvalues of $W_z$. Consider the 2D limit of the three-dimensional $\mathbb Z_N$ model where the $x$ and $y$ directions are taken to be very large compared to the $z$ direction. In this limit a non-contractible loop along the $z$-cycle becomes very small and the following perturbation is essentially local
\begin{equation}
	H = H_{3D,\mathbb Z_N} - \frac{J_z}2\left(W_z + W_z^\dagger\right),\label{eq:QuasiLocalPerturbation}
\end{equation}
where $W_z$ creates a loop along $z$.
Since this perturbation commutes with the original Hamiltonian, besides the conditions \eqref{eq:GSConditions} the ground state must also satisfy $W_z\ket{GS} = \ket{GS}$. Thus the $N^3$-fold degeneracy is not stable in the 2D limit and the $N^2$ remaining ground states are now $\ket{2D,a,b} \equiv \ket{\psi_{ab0}}$. The gap to the state $\ket{\psi_{abc}}$ is $\Delta E_c = J_c[1-\cos(\frac{2\pi }Nc)]$.

It is easy to see that $S_{yx}$ and $T_{yx}$ on this set of ground states exactly correspond the two dimensional $\mathbb Z_N$ modular matrices and can be used to construct the corresponding UMTC. Thus the 3D $\mathbb Z_N$ model and our universal quantities exactly reduce to the 2D versions in this limit. Furthermore, the 3D quasiparticle basis also directly reduce to the 2D quasiparticle basis.

\section{Quantum Double Models in Three-Dimensions}
In this section we will construct exactly soluble models in three-dimensions for any finite group $G$. These are nothing but a natural generalization of Kitaev's quantum double models \cite{Kitaev:1997wr} to three-dimensions and are closely related to discrete gauge theories with gauge group $G$. These models will have the above $\mathbb Z_N$ models as a special case, but formulated in a slightly different way.

Consider a simple cubic lattice \footnote{The model can easily be defined on arbitrary triangulations, but for simplicity we will consider the cubic lattice.} with the orientation used above. Let there be a Hilbert space $\mathcal H_l\approx\mathbb C[G]$ on each link $l$, where $G$ is a finite group, and let there be an isomorphism $\mathcal H_l\overset{\sim}{\rightarrow}\mathcal H_{l^\star}$ for the link $l$ and its reverse orientation $l^\star$ as $\ket{g_l}\mapsto\ket{g_{l^\star}}=\ket{g_l^{-1}}$. Furthermore let the natural basis of the group algebra be orthonormal. The following local operators will be useful
\begin{align*}
L^g_+\ket z &= \ket{gz}, & T^h_+\ket z &= \delta_{h,z}\ket z,\\
L^g_-\ket z &= \ket{zg^{-1}}, & T^h_-\ket z &=\delta_{h^{-1},z}\ket z.
\end{align*}
To each two dimensional plaquette $p$, associate a orientation wrt. to the lattice orientation using the right-hand rule. For such a plaquette, define the following operator
\[B_h(p)\BLvert\oneSquare{z_L}{z_U}{z_R}{z_D}{p}{0}\Brangle = \delta_{z_Uz_R^{-1}z_D^{-1}z_L,h}\BLvert\oneSquare{z_L}{z_U}{z_R}{z_D}{p}{0}\Brangle,\]
and similar for other orientations of plaquettes. Note that the order of the product is important for non-Abelian groups. To each lattice site $s$, define the operator
\[	A_g(s) = \prod_{l_-} L^g_-(l_-)\prod_{l_+} L^g_+(l_+),	\]
where $l_-$ are the set of links pointing into $s$ while $l_+$ are the links pointing away from $s$. In particular we have that
\[A_g(s) \BLvert\oneVertex{x_1}{z_2}{y_2}{x_2}{z_1}{y_1}\Brangle	= \BLvert\oneVertex{x_1g^{-1}}{gz_2}{gy_2}{gx_2}{z_1g^{-1}}{y_1g^{-1}}\Brangle.\]
From these we have two important operators
\[	A(s) = \frac 1{|G|}\sum_{g\in G}A_g(s),	\]
and $B(p)\equiv B_1(p)$, where $1\in G$ is the identity element. One can show that both these operators are hermitian projectors. Furthermore one can check that they all commute together
\begin{align*}
	\big[A(s),B(p)\big] &= 0,& \quad&\forall s,p,\\
	\big[B(p),B(p')\big] &= 0,& \quad&\forall p,p',\\
	\big[A(s),A(s')\big] &= 0,& \quad&\forall s,s'.\\ 
\end{align*}
We can now define the Hamiltonian of the three-dimensional quantum double model as
\begin{equation}
	H = -J_e\sum_sA(s) - J_m\sum_pB(p).
\end{equation}
Since the Hamiltonian is just a sum of commuting projectors, the ground states of the system must satisfy
\[	A(s)\ket{GS} = B(p)\ket{GS} = \ket{GS},	\]
for all $s$ and $p$. The ground state can be constructed using the following hermitian projector $\rho_{GS} = \prod_s A(s)\prod_p B(p)$. If we take as reference state $\ket 1 = \ket{1_{l_1} 1_{l_2}\dots}$, we can write
\[	\ket{GS} = \rho_{GS}\ket 1 = \prod_s A(s)\ket 1.	\]

\subsection{Ground states on $T^3$}

The easiest way to construct the ground states on the three-torus is to consider the minimal torus, which is just a single cube where the boundaries are identified.
The minimal torus has one site $s$
\[	\oneVertex{b}{a}{c}{b}{a}{c}	\]
and three plaquettes $p_1$, $p_2$, $p_3$
\[\oneSquare{a}{b}{a}{b}{p_1}{0} \quad \oneSquare{a}{c}{a}{c}{p_2}{1}\quad \oneSquare{c}{b}{a}{c}{p_3}{1}\]
One can readily show that the subspace $\mathcal H^{B=1}$ satisfying $B(p)\ket{GS} \overset != \ket{GS}$ for $p=p_1,p_2,p_3$, is spanned by the vectors $\ket{a,b,c}$ such that $ab=ba$, $bc = cb$ and $ac=ca$. The last condition is $A(s)\ket{GS}=\ket{GS}$ where on the basis vectors
\[	A(s)\ket{a,b,c} = \frac 1{|G|}\sum_{g\in G}\ket{gag^{-1},gbg^{-1},gcg^{-1}}.\]
In the case of Abelian groups $G$, this condition is clearly trivial and then we have $GSD=|G|^3$. In general we can find the ground state degeneracy by taking the trace of the projector $A(s)$ in $\mathcal H^{B=1}$. This is given by
\begin{align*}
	GSD&=\sum_{\{a,b,c\}}\bra{a,b,c}A(s)\ket{a,b,c}\\ &= \frac 1{|G|}\sum_{g\in G}\sum_{\{a,b,c\}}\delta_{ag,ga}\delta_{bg,gb}\delta_{cg,gc},
\end{align*}
where $\{a,b,c\}$ is triplets of commuting group elements.
One can actually easily check that the following vectors span the ground state subspace
\begin{equation}
	\ket{\psi_{[a,b,c]}} = \frac 1{|G|}\sum_{g\in G}\ket{gag^{-1},gbg^{-1},gcg^{-1}},
	\end{equation}
where $[a,b,c]=\{(\tilde a,\tilde b,\tilde c)\in G\times G\times G\,|\,(\tilde a,\tilde b,\tilde c)=(gag^{-1},gbg^{-1},gcg^{-1}),g\in G\}$ is the three-element conjugacy class and $a,b,c$ are representatives of the class.

\subsection{$3D$ $\tilde S$ and $\tilde T$ matrices and the $SL(2,\mathbb Z)$ subgroup}
We can now readily compute the overlaps \eqref{eq:overlap} for the above model for any group $G$. We find the following representations of $\texttt{MCG}(T^3)=SL(3,\mathbb Z)$
\[	\tilde S_{[a,b,c],[\bar a,\bar b,\bar c]} = \bra{\psi_{[a,b,c]}}\:\tilde S\:\ket{\psi_{[\bar a,\bar b,\bar c]}} = \delta_{[a,b,c],[\bar b,\bar c,\bar a]}	\]
and
\[	\tilde T_{[a,b,c],[\bar a,\bar b,\bar c]} = \bra{\psi_{[a,b,c]}}\:\tilde T\:\ket{\psi_{[\bar a,\bar b,\bar c]}} = \delta_{[a,b,c],[\bar a,\bar a\bar b,\bar c]},	\]
since $\tilde S\ket{\psi_{[a,b,c]}}=\ket{\psi_{[b,c,a]}}$ and $\tilde T\ket{\psi_{[a,b,c]}}=\ket{\psi_{[a,ab,c]}}$.

Once again we can consider the subgroup $SL(2,\mathbb Z)\subset SL(3,\mathbb Z)$ generated by \eqref{eq:SL2ZsubgroupGenerators}. The representation of this subgroup can be directly computed and is given by
\[	S^{yx}_{[a,b,c],[\bar a,\bar b,\bar c]} = \bra{\psi_{[a,b,c]}}\:S^{yx}\:\ket{\psi_{[\bar a,\bar b,\bar c]}} = \delta_{[a,b,c],[\bar b,\bar a^{-1},\bar c]}	\]
and
\[	T^{yx}_{[a,b,c],[\bar a,\bar b,\bar c]} = \bra{\psi_{[a,b,c]}}\:T^{yx}\:\ket{\psi_{[\bar a,\bar b,\bar c]}} = \delta_{[a,b,c],[\bar a,\bar a\bar b,\bar c]}.	\]
Note that since $c$ is not independent of $a$ and $b$, in general we don't have the decomposition $S^{3D}_G=\bigoplus_{n=1}^{|G|}S^{2D}_G$ and $T^{3D}_G=\bigoplus_{n=1}^{|G|}T^{2D}_G$, unless the group is Abelian.

\subsection{Branching Rules and Dimensional Reduction}
With the above formulas, we can directly compute the $\tilde S$ and $\tilde T$ generators for any group G. In the limit where one direction of the 3-torus is taken to be very small, we can view the 3D topological order as several 2D topological orders.

The branching rules \eq{C3C2} for the dimensional reduction can be directly
computed by studying how a representation of $SL(3,\mathbb Z)$ decomposes into
representations of the subgroup $SL(2,\mathbb Z)\subset SL(3,\mathbb Z)$. For
example, for some of the simplest non-Abelian finite groups we find the branching rules
\begin{gather*}
	\cC^{3D}_{S_3} = \cC^{2D}_{S_3}\oplus\cC^{2D}_{\mathbb Z_3}\oplus \cC^{2D}_{\mathbb Z_2},\\
	\cC^{3D}_{D_4} = 2\,\cC^{2D}_{D_4}\oplus 2\,\cC^{2D}_{D_2}\oplus \cC^{2D}_{\mathbb Z_4},\\
	\cC^{3D}_{D_5} = \cC^{2D}_{D_5}\oplus 2\,\cC^{2D}_{\mathbb Z_5}\oplus \cC^{2D}_{\mathbb Z_2},\\
	\cC^{3D}_{S_4} = \cC^{2D}_{S_4}\oplus \cC^{2D}_{D_4}\oplus \cC^{2D}_{D_2}\oplus \cC^{2D}_{\mathbb Z_4}\oplus \cC^{2D}_{\mathbb Z_3}.
\end{gather*}
In general we find the following branching in the dimensional reduction
$\cC_G^{3D} = \bigoplus_{C}\cC_{G_C}^{2D}$,
where $\bigoplus_{C}$ sums over all different conjugacy classes $C$ of $G$, and
$G_C$ is the centralizer subgroup of $G$ for some representative $g_C\in C$.
Similar to the $G=\mathbb Z_N$ case above \eqref{eq:QuasiLocalPerturbation},
the degeneracy between the different sectors can be lifted by a perturbation
creating Wilson loops along the small non-contractible cycle of $T^3$, which is essentially a local
perturbation in the 2D limit.

We like to remark that the above branching result for dimensional reduction can
be understood from a ``gauge symmetry breaking'' point of view.  In the
dimensional reduction, we can choose to insert gauge flux through the
small compactified circle. The different choices of the gauge flux is given by
the conjugacy classes $C$ of $G$.  Such gauge flux break the ``gauge symmetry''
from $G$ to $G_C$. So, such a compactification leads to a 2D gauge theory with
gauge group $G_C$ and reduces the 3D topological order $\cC^{3D}_G$ to a 2D
topological order $\cC^{2D}_{G_C}$.  The different choices of gauge flux lead to
different degenerate 2D topological ordered states, each described by
$\cC^{2D}_{G_C}$ for a certain $G_C$.  This gives us the result \eqn{G3G2}.  It
is quite interesting to see that the branching \eq{R3R2} of the representation
of the mapping class group $SL(3,\mathbb Z) \to SL(2,\mathbb Z)$ is closely
related to the ``gauge symmetry breaking'' in our examples.

In order to gain a better understanding of the information contained in these
branching rules, we will consider a simple example. 

\begin{table*}[!ht]
\centering
\begin{tabular}{c"c|c|c|c|c|c|c|c}
$\otimes$&$\bm 1$& $A^1$ &$A^2$ 					 &$B$ 										   &$B^1$ 							  			 &$C$					   &$C^1$ & $C^2$  \\
\thickhline
$ \bm 1 $&$\bm 1$&$A^1$  &$A^2$						 &$B$ 										   &$B^1$   									 &$C$					   &$C^1$	           & $C^2$ \\
$A^1$	 &$A^1$	 &$\bm 1$&$A^2$						 &$B^1$	      							   	   &$B$											 &$C$                      &$C^1$ &$C^2$\\
$A^2$ 	 &$A^2$	 &$A^2$  &$\bm 1\oplus A^1\oplus A^2$&$B\oplus B^1$				 				   &$B\oplus B^1$								 &$C^1\oplus C^2$          &$C\oplus C^2$&$C\oplus C^1$\\
$B$		 &$B$    &$B^1$  &$B\oplus B^1$				 &$\bm 1\oplus A^2\oplus C\oplus C^1\oplus C^2$&$A^1 \oplus A^2\oplus C\oplus C^1\oplus C^2$ &$B\oplus B^1$ 		   &$B\oplus B^1$ &$B\oplus B^1$\\
$B^1$ 	 &$B^1$	 &$B$ 	 &$B\oplus B^1$				 &$A^1 \oplus A^2\oplus C\oplus C^1\oplus C^2$ &$\bm 1\oplus A^2\oplus C\oplus C^1\oplus C^2$&$B\oplus B^1$ 		   &$B\oplus B^1$ &$B\oplus B^1$\\
$C$ 	 &$C$ 	 &$C$    &$C^1\oplus C^2$			 &$B\oplus B^1$						 		   &$B\oplus B^1$				 				 &$\bm 1\oplus A^1\oplus C$&$C^2\oplus A^2$ &$C^1\oplus A^2$\\
$C^1$ 	 &$C^1$  &$C^1$  &$C\oplus C^2$				 &$B\oplus B^1$	 						       &$B\oplus B^1$								 &$C^2\oplus A^2$          &$\bm 1\oplus A^1\oplus C^1$ &$C\oplus A^2$\\
$C^2$ 	 &$C^2$	 &$C^2$  &$C\oplus C^1$				 &$B\oplus B^1$								   &$B\oplus B^1$								 &$C^1\oplus A^2$          &$C\oplus A^2$ &$\bm 1\oplus A^1\oplus C^2$
\end{tabular}\caption{Fusion rules of two-dimensional $D(S_3)$ model. Here $B$ and $C$ correspond to pure flux excitations, $A^1$ and $A^2$ pure charge excitations, $\bm 1$ the vacuum sector while $B^1$, $C^1$ and $C^2$ are charge-flux composites.}\label{tab:2DS3FusionTules}
\end{table*}

\section{Example: $G=S_3$}

\subsection{Two-Dimensional $D(S_3)$}

Let us consider the simplest non-Abelian group $G=S_3$.
Let us first recall the 2D quantum double models. The excitations of these models are given by irreducible representations of the Drinfeld Quantum Double $D(G)$. The states can be labelled by $\ket{C,\rho}$, where $C$ denote a conjugacy class of $G$ while $\rho$ is a representation of the centralizer subgroup $G_C \equiv Z(a)=\{g\in G|ag=ga\}$ of some element in $a\in C$ (note that $Z(a)\approx Z(gag^{-1})$).

The symmetric group $G=S_3$ consists of the elements $\{(),(23),(12),(123),
(132), (13) \}$, where $(\dots)$ is the standard notation for cycles (cyclic
permutations). There are three conjugacy classes $A =\{()\} $, $B =
\{(12),(13),(23)\}$ and $C = \{(123),(132)\}$, with the corresponding
centralizer subgroups $G_A=S_3$, $G_B=\mathbb Z_2$, $G_C=\mathbb Z_3$. The
number of irreducible representations for each group is equal to the number of
conjugacy classes, $3$ for $G_A$ and $G_C$ while $2$ for $G_B$. For simplicity we will label the particles corresponding to
the three different conjugacy classes by $(\bm 1, A^1,A^2)$, $(B,B^1)$ and
$(C,C^1,C^2)$. Here the particles without a superscript, $B$ and $C$, are pure fluxes
(trivial representation), $A^1$ and $A^2$ are pure charges (trivial
conjugacy class), while $B^1$, $C^1$ and $C^2$ are charge-flux composites. The fusion rules for the two-dimensional $D(S_3)$ model is
given in table \ref{tab:2DS3FusionTules}.

\subsection{Three-Dimensional $G=S_3$ Model}

In three dimensions, the $S_3$ model has two point-like topological
excitations, which are pure charge excitations that can be labelled by
$A^1_{3D}$ and $A^2_{3D}$.  Here $A^1$ is the one-dimensional irreducible
representation of $S_3$ and $A^2$ the two-dimensional irreducible
representation of $S_3$.  Under the dimensional reduction to 2D, they become the
2D charge particles labelled by $A^1$ and $A^2$.  The $S_3$ model also has two
string-like topological excitations, labelled by the non-trivial conjugacy
classes $B_{3D}$ and $C_{3D}$.  Under the dimensional reduction to 2D, they
become the 2D particles with pure fluxes described by $B$ and $C$.  (For
details, see the discussion below.)  We can also add a 3D charged particle to a
3D string and obtain a so called mixed string-charge excitation. Those
mixed string-charge excitations are labelled by $B^1_{3D}$, $C^2_{3D}$, and
$C^3_{3D}$, and, under the dimensional reduction, become the 2D particles $B^1$,
$C^2$, and $C^3$.

We like to remark that, since a 3D string carries gauge flux described by a
conjugacy class $B$ or $C$, the $S_3$ ``gauge symmetry'' is broken down to
$G_B=\mathbb Z_2$ on the $B_{3D}$ string, and down to $G_C=\mathbb Z_3$  on the $C_{3D}$
string.

Under the symmetry breaking $S_3 \to \mathbb Z_2$, the two irreducible representations $A^1$ and $A^2$ of $S_3$ reduce to the irreducible
representations $1$ and $e$ of $\mathbb Z_2$: $A^1 \to e$ and $A^2 \to 1\oplus e$.  Thus
fusing the $S_3$ charge $A^1_{3D}$ to a $B_{3D}$ string give us the mixed
string-charge excitation $B^1_{3D}$.  But fusing the $S_3$ charge $A^2_{3D}$ to
a $B_{3D}$ string gives us a composite mixed string-charge excitation
$B_{3D}\oplus B^1_{3D}$. (The physical meaning of the composite topological
excitations  $B_{3D}\oplus B^1_{3D}$ is explained in \Ref{LW1384}.) So fusing the two
non-trivial $S_3$ charges to a $B_{3D}$ string only give us one mixed
string-charge excitation $B^1_{3D}$.  

Under the symmetry breaking $S_3 \to \mathbb Z_3$, the two irreducible representations
$A^1$ and $A^2$ of $S_3$ reduce to the irreducible representations $1$, $e_1$
and $e_2$ of $\mathbb Z_3$: $A^1 \to 1$ and $A^2 \to e_1\oplus e_2$.  Thus fusing the $S_3$
charge $A^1$ to a $C_{3D}$ string still gives us the string excitation
$C_{3D}$.  But fusing the $S_3$ charge $A^2_{3D}$ to a $C_{3D}$ string gives us
a composite mixed string-charge excitation $C^1_{3D}\oplus C^2_{3D}$.  So fusing the two non-trivial $S_3$ charges to a $C$ string give
us two mixed string-charge excitations $C^1_{3D}$ and $C^2_{3D}$.  We see that
the fusion between point $S_3$ charges and the strings is consistent with
fusion of the corresponding 2D particles.

Now, we would like to understand the fusion and braiding properties of the 3D
strings $B_{3D}$ and $C_{3D}$.  To do that, let us consider the dimension
reduction $ \cC^{3D}_{S_3} = \cC^{2D}_{S_3}\oplus\cC^{2D}_{\mathbb Z_3}\oplus
\cC^{2D}_{\mathbb Z_2}$.  Let us choose the gauge flux through the small
compactified circle to be $B$.  In this case $\cC^{3D}_{S_3} \to
\cC^{2D}_{\mathbb Z_2}$.  $\cC^{2D}_{\mathbb Z_2}$ is a $\mathbb Z_2$ topological order
in 2D and contains four particle-like topological excitations $\bm 1$, $e,\ m,
f$, where $\bm 1$ is the trivial excitations.  $e$ is the $\mathbb Z_2$ charge and $m$
the $\mathbb Z_2$ vortex, which are both bosons.  $f$ is the bound state of $e$ and $m$
which is a fermion.  The trivial 2D excitation $\bm 1$ comes from the trivial
3D excitation $\bm 1_{3D}$, and the $\mathbb Z_2$ charge $e$ comes from the 3D charge
excitation $A^1$.  The 3D string excitations $B$ and $B^1$, wrapping around the
small compactified circle, give rise to two particle-like excitations in 2D --
the $\mathbb Z_2$ vortex $m$ and the fermion $f$.  In the dimensional reduction, the
gauge flux $B$ through the small compactified circle forbids the 3D string
excitations $C_{3D}$, $C^1_{3D}$, and $C^2_{3D}$ to wrap around the small
compactified circle. So there is no 2D excitations that correspond to the 3D
string excitations $C_{3D}$, $C^1_{3D}$, and $C^2_{3D}$.  Because of the
symmetry breaking $S_3 \to \mathbb Z_2$ caused by the gauge flux $B$, the 3D particle
$A^2_{3D}$ reduces to $\bm 1\oplus e$ in 2D.  

\begin{figure}[tb]
\centering
\includegraphics[width=.15\textwidth]{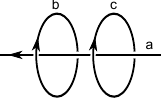}
\caption{Three string configuration, where two loops of type $b$ and 	$c$ are threaded by a string of type $a$.}\label{fig:threeString}
\end{figure}

The above results have a 3D understanding.  Let us consider the situation where
two loops, $b$ and $c$, are threaded by string $a$ (see Fig.
\ref{fig:threeString}). If the $a$-string is the type-$B_{3D}$ string, then the
$b$ and $c$-strings must also be the type-$B_{3D}$ string.  So the type
$B_{3D}$ string in the center forbids the 3D strings $C_{3D}$, $C^1_{3D}$, and
$C^2_{3D}$ to loop around it.  This is just like the gauge flux $B$ through
the small compactified circle forbids the 3D string excitations $C_{3D}$,
$C^1_{3D}$, and $C^2_{3D}$ to wrap around the small compactified circle.  So the type-$B_{3D}$ string in the center corresponds to the gauge flux $B$
through the small compactified circle.

The fusion and braiding of the 2D particle $e$ is very simple: it is an boson
with fusion $e\otimes e =\bm 1$.  This is consistent with the fact that the
corresponding 3D particle $A^1_{3D}$ is a boson with fusion $A^1_{3D}\otimes A^1_{3D} =\bm 1_{3D}$.  The fusion and braiding of the 2D particle $m$ is also
very simple, since it is also an boson $m\otimes m=\bm 1$.  This suggests that
the 3D type-$B_{3D}$ string excitations has a simple fusion and braiding
property, \emph{provided that those  3D string excitations are threaded by a
type-$B_{3D}$ string going through their center} (see Fig.
\ref{fig:threeString}).  For example, from the 2D fusion rule $m\otimes m=\bm
1$, we find that the fusion of two type-$B_{3D}$ loops give rise to a trivial
string 
\begin{align}
\label{BBZ2}
B_{3D}\otimes B_{3D}=\bm 1_{3D}.
\end{align}
As suggested by the 2D braiding of two $m$ particles, when a type-$B_{3D}$
string going around another type-$B_{3D}$ string, the induced phase is zero
(\ie the mutual braiding ``statistics'' is trivial).

Similarly,  we can choose the gauge flux through the small compactified circle
to be $C$.  In this case $\cC^{3D}_{S_3} \to \cC^{2D}_{\mathbb Z_3}$, and
$\cC^{2D}_{\mathbb Z_3}$ is a $\mathbb Z_3$ topological order in 2D which has 9
particle types: $\bm 1$, $e_1$, $e_2$, $m_1$, $m_2$, $e_im_j|_{i,j=1,2}$.  In
this case, the gauge flux $C$ through the small compactified circle forbids the
3D string excitations $B_{3D}$ and $B^1_{3D}$ to wrap around the small
compactified circle. So there is no 2D excitations that correspond to the 3D
string excitations $B_{3D}$ and $B^1_{3D}$.  The 3D string excitation $C_{3D}$
wrapping around the small compactified circle gives rise to a composite $\mathbb Z_3$
vortex $m_1\oplus m_2$ in 2D. (This is because there are two non-trivial group
elements in $S_3$ that commute with a group element in the conjugacy class
$C$). Also, from the $S_3 \to \mathbb Z_3$ symmetry breaking: $A^1 \to 1$ and $A^2 \to
e_1\oplus e_2$, we see that the 3D $A^1_{3D}$ charge reduces to type-$\bm 1$
particle in 2D, and the 3D $A^2_{3D}$ charge reduce to a composite particle
$e_1\oplus e_2$ in 2D.
 
The fusion of the composite 2D particle $c=m_1\oplus m_2$ is given by
\begin{align}
 c\otimes c = 2 \bm 1\oplus c .
\end{align}
This leads to the corresponding fusion rule for the 3D type-$C_{3D}$ loops 
\begin{align}
\label{CCZ3}
 C_{3D}\otimes C_{3D} = 2 \bm 1_{3D}\oplus C_{3D} \text{ or }  \bm 1_{3D}\oplus A^1_{3D}\oplus C_{3D},
\end{align}
\emph{provided that those  3D loops are threaded by a type-$C_{3D}$ string going
through their center} (see Fig. \ref{fig:threeString}).  (The ambiguity arises
because the 3D charge $A^1_{3D}$ reduces to $\bm 1$ in 2D.)

Now, let us choose the gauge flux through the small compactified circle to be
trivial.  In this case $\cC^{3D}_{S_3} \to \cC^{2D}_{S_3}$, which has 8
particle types: $\bm 1$, $A^1$, $A^2$, $B$, $B^1$, $C$, $C^1$, $C^2$.  The 3D
string excitation $B_{3D}$ and $C_{3D}$ wrapping around the small compactified
circle gives rise to the 2D excitation $B$ and $C$.  The fusion of the 2D
particle $C$ is given by
\begin{align}
 C\otimes C = \bm 1\oplus A^1\oplus C.
\end{align}
This leads to the corresponding fusion rule for the 3D type-$C_{3D}$ loops 
\begin{align}
\label{CC}
 C_{3D}\otimes C_{3D} = \bm 1_{3D}\oplus A^1_{3D}\oplus C_{3D},
\end{align}
\emph{provided that those 3D loops are not threaded by any non-trivial
string}. The above fusion rule implies that when we fusion
two $C_{3D}$ loops, we obtain three accidentally degenerate states:
the first one is a non-topological excitation,
the second one is a $S_3$ charge $A^1_{3D}$, and the third one
is a  $S_3$ string $C_{3D}$.

Similarly, the fusion of the 2D particle $B$ is given by
\begin{align}
 B\otimes B = \bm 1\oplus A^2\oplus C\oplus C^1\oplus C^2.
\end{align}
This leads to the corresponding fusion rule for the 3D type-$B_{3D}$ loops 
\begin{align}
\label{BB}
 B_{3D}\otimes B_{3D} = \bm 1_{3D}\oplus A^2_{3D}\oplus C_{3D}\oplus C^1_{3D}\oplus C^2_{3D}.
\end{align}
This way, we can obtain the fusion algebra between all the 3D excitations
$A^1_{3D}$, $A^2_{3D}$, $B_{3D}$, $B^1_{3D}$, $C_{3D}$, $C^1_{3D}$, $C^2_{3D}$.

\begin{table}[t]
\centering
\begin{tabular}{c"c|c|c}
$a$&$A$&$B$& $C$   \\
\thickhline
Symmetry Breaking	& $S_3\rightarrow S_3$ & $S_3\rightarrow \mathbb Z_2$	& $S_3\rightarrow \mathbb Z_3$\\
$\bm 1_{3D}\rightarrow$ &	$\bm 1$	& $\bm 1$	& $\bm 1$	\\
$A_{3D}^1\rightarrow$	  & $A^1$	& $e$	& $\bm 1$	\\
$A_{3D}^2\rightarrow$	  &	$A^2$	& $\bm 1\oplus e$	& $e_1\oplus e_2$	\\
$B_{3D}\rightarrow$	  	  &	$B$		&	m	&	-\\
$B_{3D}^1\rightarrow$	  &	$B^1$	&	em	&	-\\
$C_{3D}\rightarrow$	  	  &	$C$		& -	& $m_1\oplus m_2$	\\
$C_{3D}^1\rightarrow$	  &	$C^1$	& -	& $e_1m_1\oplus e_1m_2$	\\
$C_{3D}^2\rightarrow$	  &	$C^2$	& -	& $e_2m_1\oplus e_2m_2$	\\
\end{tabular}\caption{The situation of figure \ref{fig:threeString}, where strings are wrapped around another string of type $a=A,B,C$. Depending on $a$, fusion algebra and braiding statistics of each string will be related to a particle of some 2D topological order, as computed from the branching rules \eqref{G3G2}. See the text for more details. }\label{tab:SymmetryBreaking}
\end{table}

On the other hand, since the above 3D string loops are not threaded by any
non-trivial string, we can shrink a single loop into a point.  So we
should be able to compute the fusion of 3D loops by shrinking them into a
points. Mathematically we will define shrinking operation $\mathcal S$, which describes the shrinking process of loops. 

Let $\mathcal E$ denote the set of 3D particle and string excitations. We would like to make sure that the shrinking operation is consistent with the fusion rules, ie $\mathcal S(a\otimes
b)=\mathcal S(a)\otimes\mathcal S(b)$ for $a,b\in\mathcal E$. One can indeed check that this is the case for the following shrinking operations
\begin{gather*}
	\mathcal S(C_{3D})=\bm 1_{3D}\oplus A^1_{3D}, \ \ \ \mathcal S(C^1_{3D})=A^2_{3D},\ \ \ \mathcal S(C^2_{3D})=A^2_{3D},\\
	\mathcal S(B_{3D})=\bm 1_{3D}\oplus A^2_{3D}, \qquad \mathcal S(B^1_{3D})=A^1_{3D}\oplus A^2_{3D}. 
\end{gather*}
So indeed, we can
compute the fusion of 3D loops by shrinking them into points.  In particular,
we find that the topological degeneracy for $N$ type-$C_{3D}$ loops is $2^N/2$.
The topological degeneracy for two type-$B_{3D}$ loops is $2$.  The topological
degeneracy for $N$ type-$B_{3D}$ loops is of order $3^N$ in large $N$ limit.

The above example suggests the following. Given a topological order in 3D,
$\cC^{3D}$, one may want to consider the situation illustrated in figure
\ref{fig:threeString} where two loops $b$ and $c$ are threaded with a string
$a$, and ask about the three-string braiding statistics. One way to compute
this is to put the system on a 3-torus and compute the quantities
\eqref{eq:overlap}, which give rise to a $SL(3,\mathbb Z)$ representation. Then
by finding the branching rules of this representation wrt. to the subgroup
$SL(2,\mathbb Z)\subset SL(3\,\mathbb Z)$, one finds how the systems decomposes
in the 2D limit $\cC^{3D}=\bigoplus_i\cC^{2D}_i$, where there will be a sector
$i$ for each string type. The three-string statistics with string $a$ in the
middle, will be related to the 2D topological order $\cC^{2D}_a$.  
To summarize:
\begin{itemize}
\item
The representation branching rule \eq{R3R2} for $SL(3,\mathbb Z)\to
SL(2,\mathbb Z)$ leads to the dimension reduction branching rule \eq{C3C2}.
\item
The number of the $SL(2,\mathbb Z)$ representations (or the number of induced
2D topological orders) is equal to the number of 3D string types in the 3D
topological order $\cC^{3D}$.
\item
The  $SL(2,\mathbb Z)$ representations also contains information about
two-string/three-string fusion, as described by eqns.
(\ref{BBZ2},\ref{CCZ3},\ref{CC},\ref{BB}).  The two-string/three-string
braiding can be obtained directly from the correspond 2D braiding of the
corresponding particles.
\end{itemize}

%

\section{Some general considerations}

To calculate the braiding statistics of strings and particles, we first need to
know the topological degeneracy $D$ in the presence of strings and particles
before they braid. This is because the unitary matrix that describe the
braiding is $D$ by $D$ matrix.  To compute the topological degeneracy $D$, we
need to know the topological types of strings and the particles since the
topological degeneracy $D$ depends on those types.

We have seen that, from the branching rules of $SL(3,\Z)$ representation under
$SL(3,\Z)\to SL(2,\Z)$ (see \eqn{R3R2}) we can obtain the number of the string
types.  How to obtain the number of the particle types?

To compute the number of the particle types, we start with a 3D sphere $S^3$,
and then remove two small balls from it.  The remaining 3D sphere will have two
$S^2$ surfaces.  This two surfaces may surround a particle and anti-particle. So
the number of the particle types can be obtained by calculating the ground state
degeneracy.  But there is one problem with this approach, the  two surfaces may
carry gapless boundary excitations or some irrelevant symmetry breaking states.

To fix this problem, we note that the 3D space $S^2\times I$ also have have two
$S^2$ surfaces, where $I$ is the 1D segment: $I=[0,1]$.  We can glue the space
$S^2\times I$ onto the 3D sphere $S^3$ with two balls removed, along the two 2D
spheres $S^2$.  The resulting space is $S^2\times S^1$.  This way, we show that
the topological degeneracy on $S^2\times S^1$ is equal to the number of the
particle types.  

For the gauge theory of finite gauge group $G$, the topologically degenerate
ground states on $S^2\times S^1$ are labelled by the group elements $g\in G$
(which describe the monodromy along the non-contractible loop in $S^2\times
S^1$), but not in an one-to-one fashion.  Two elements $g$ and $g'=h^{-1} g h$
label the same ground state since $g$ and $g'$ are related by a gauge
transformation.  So the topological degeneracy on $S^2\times S^1$ is equal to
the number of conjugacy classes of $G$.  The number of conjugacy classes is
equal to the number of irreducible representations of $G$, which is also the
number of the particle types, a well known result for gauge theory.

Once we know the types of particles and strings, the simple fusion and braiding
of those excitations can be obtained from the dimensional reduction as
described in this paper.

\section{Conclusion}

In a recent work \Ref{MW14}, we proposed that for a gapped $d$-dimensional
theory on a manifold $\mathcal M$, the overlaps \eqref{eq:overlap} give rise to
a representation of $\texttt{MCG}(\mathcal M)$ and that these are robust
against any local perturbation that do not close the energy gap. In this paper
we studied a simple class of $\mathbb Z_N$ models on $\mathcal M=T^3$ and
computed the corresponding representations of $\mathtt{MCG}(T^3)=SL(3,\mathbb
Z)$. We argued that, similar to in 2D, the $\tilde T$ generator contains
information about particle and string excitations above the ground state,
although computed from the ground states. In an independent work \Ref{JMR1462},
the authors studied the matrices \eqref{eq:overlap} using some Abelian models
on $T^3$. They argued that the generator $\tilde S$ contains information about
braiding processes involving three loops.

Furthermore we studied a dimensional reduction process in which the 3D
topological order can be viewed as several 2D topological orders
$\cC^{3D}=\bigoplus_i\cC^{2D}_i$. This decomposition can be computed from
branching rules of a $SL(3,\mathbb Z)$ representation into representations of a
$SL(2,\mathbb Z)\subset SL(3,\mathbb Z)$ subgroup. Interestingly, this
reduction encodes all the information about three-string statistics discussed
in \Ref{WL1437} for Abelian groups. This approach, however, also
provide information about fusion and braiding statistics of non-Abelian string
excitations in 3D.

We also discussed how to obtain information about particles by putting the theory
on $S^2\times S^1$. All this lends support for our conjecture\cite{MW14}, that the overlaps
\eqref{eq:overlap} for different manifold topologies $\mathcal M$, completely characterize
topological order with finite ground state degeneracy in any dimension.

\vspace{10 mm}
This research is supported by NSF Grant No.
DMR-1005541, NSFC 11074140, and NSFC 11274192. It is also supported by the John
Templeton Foundation. Research at Perimeter Institute is supported by the
Government of Canada through Industry Canada and by the Province of Ontario
through the Ministry of Research.

\bibliography{bib/wencross,bib/all,bib/publst,bib/local}

\newpage
\appendix




\end{document}